# Optical 3D-storage in sol-gel materials
# with a reading by Optical Coherence Tomography-technique.

Jorge-Alejandro Reyes-Esqueda[a,1], Laurent Vabre[b,2],

Romain Lecaque[b], François Ramaz[b], Benoît C. Forget[b], Arnaud Dubois[b], Bernard Briat[b], Claude Boccara[b],

Gisèle Roger[a], Michael Canva[a], Yves Lévy[a],

Frédéric Chaput[c], Jean-Pierre Boilot[c].

[a] Groupe d'Optique Non-linéaire, Laboratoire Charles Fabry de l'Institut d'Optique Théorique et Appliquée, CNRS UMR 8501, Ecole Supérieure d'Optique, Bât 503, Université Paris Sud, 91403, Orsay Cedex, France.

[b] Laboratoire d'Optique Physique, CNRS UPR A0005, UPMC, Ecole Supérieure de Physique et de Chimie Industrielles, 10 rue Vauquelin, 75005 Paris, France.

[c] Groupe de Chimie du Solide, Laboratoire de Physique de la Matière Condensée, CNRS UMR 7643, Ecole Polytechnique, 91128 Palaiseau Cedex, France.

**Abstract**

We report on the recording of 3D optical memories in sol-gel materials by using a non-linear absorption effect. This effect induces a local change of the optical properties of the material which is read and quantified with a high resolution full-field Optical Coherence Tomography setup. It is the first time that this technique is used for this purpose. Data recording was performed by focused picosecond (ps) single-pulse irradiation at 1064 nm with energy densities of 10 and 33 J/cm$^2$ per pulse.

*PACS*: 42.70.M; 42.30.W; 42.70.L; 42.65.

*Key words*: Non-linear Materials; Sol-gel; Optical Coherence Tomography; Optical Data Storage; Non-linear Absorption; Two Photon Absorption.

[1] Supported by Conacyt (Mexico) and the Ministère de la Recherche (France).
[2] Supported by the Délégation Générale pour l'Armement (France).

## 1. Introduction

In recent years, several strong efforts have been made in order to increase storage computer capability as the limit of the conventional media ($2.5 \times 10^9$ bits/cm$^2$ read at $\lambda=200$ nm[1]) has been almost reached. The removability, replicability, durability, lightness, and its low cost have made optical memory more advantageous than magnetic data storage. However, there are some unbreakable physical limits for this kind of memory. One of these barriers, the diffraction of laser light, limits the density of stored information to the reciprocal of the wavelength raised to the power of the dimensions used to store information ($\rho \sim 1/\lambda^2$ in two-dimensional systems, where $\rho$ is the density of information and $\lambda$ the wavelength). This equation suggests that the information storage is much higher for UV than for IR light, nevertheless the shorter the wavelength, the stronger the Rayleigh scattering inside the medium. However, wavelength is not the only parameter that determines the stored data density although it remains an important one.

Remaining within these borders, the recently available digital versatile disk (DVD) system uses a laser with a shorter wavelength than for compact disc (CD) and also a lens with a larger numerical aperture, which allows a rise of the density storage by a factor of 7.3[2]. In order to be able to fulfill the increasing demand in the information storage field, several methods to break those density limits have been tried. Hybrid near-field optics[3] and other three-dimensional (3D) optical recording techniques like optical spectral hole burning[4], holographic data storage[5] and two photon absorption (TPA) [1, 6-14] have been recently used in this direction.

In the case of multi-photon absorption, there exists a threshold above which a change of the optical properties into a reduced volume may be induced. This change occurs at the focal point of a laser beam. This fact ensures avoiding cross talk in the writing phase and since TPA depends on the square of the radiation flux intensity, the optical properties will change within a localized volume in the range of $\lambda^3$. For an IR laser, an interaction volume of 1 µm$^3$ corresponds to a storage density of 1 Tbit/cm$^3$. According to the materials' properties, the main physical mechanisms used for recording by TPA are photopolymerization[11], photochromicity[7], dimer-monomer transformation[1], control of the media acidity[1], diffusional redistribution of fluorescent molecules[12] and photobleaching[3, 13, 14]. Another important recording mechanism indirectly linked to TPA because of the multi-photon feature of the absorption, is the optical damaging of a material by micro-explosion[15, 16] in the region of the focus of laser irradiation.

In the present work, we write binary data by using a focused single laser shot at 1064 nm. The writing mechanism is related to TPA and micro-explosion processes, as it will be shown below. So far, crystals[17], silica glasses[15], and more often polymers[3, 5, 6] have been used for recording 3D structures. To the best of our knowledge, it is the first time that sol-gel materials, either undoped or doped with organic dyes, are being used as storage media.

Considering the reading process of the data, essentially single-[18], two-[19] and multi-photon[20] techniques have been used. These are mainly based on the measurement of fluorescence or transmission from the memory. In a few cases, data have been read through differential interference contrast[11] and confocal microscopies[2]. The technique presented here is quite different and original. By means of Optical Coherence Tomography (OCT) [21], we can select the backscattered light from a particular layer inside the volume of the media, because of the small coherence length of the illuminating light source. Again, it seems to be the first time that full-field OCT is used for reading information from a 3D optical memory.

We present in this paper original results concerning optical volume storage, where binary data are recorded by using the non-linear properties of hybrid sol-gel samples doped with organic dyes and read with a new and efficient technique. In section 2, we present the materials used as well as the recording and the reading setups. Then, in section 3 we present the results and the discussion, and finally we give our conclusions.

## 2. Experimental methods

### a) Materials

The method for general synthesis of silica-based doped xerogels has been previously reported[22]. Hydrolysis of the organically modified silicon precursor was performed under acidic conditions with acetone as the common solvent. After several hours of hydrolysis at room temperature, a small amount of amine modified silicon alkoxide was added to neutralize the acidity of the medium, thereby increasing the condensation reaction rate. An acetonic solution of laser dye was then added to yield a concentration in the range from $10^{-4}$ to $10^{-3}$ mol/l. Afterward, the resulting sol was poured into polypropylene cylindrical-shaped molds and sealed. After drying at 70 °C for several weeks dense monolithic xerogels (30 mm diameter and 10 mm thick) were obtained. Finally samples were machined and polished (around 4 nm of roughness). For the purposes of this work, two $SiO_2$ sol-gel samples, one undoped and one doped, were selected to be tested as optical memories. As a doping molecule, we chose one molecule from the Perylene families, Perylene orange. As it was previously reported, this molecule is very sensitive to pulsed irradiation[23]. Moreover, previous results show that its spectroscopic features make it very suitable for TPA with picosecond pulses at 1064 nm[24].

**b) Recording**

The information storage setup is shown schematically in figure 1. A Nd:YAG laser emitting 45 picosecond (ps) pulses, at 1064 nm, with a repetition rate of 10 Hz, was used for writing information into the material. Single-shots were obtained by using a laser-synchronized shutter. Neutral optical densities were used to control the intensity of the incident light during the recording process. The energy densities used for the inscription were 0.1, 0.3, 1, 3.3, 10, 33 and 100 $J/cm^2$. Moreover, we tuned the total incident energy by controlling the number of incident shots onto the sample. In this way, 1, 2, 4, 8, 16 and 32 single-shots were used to write each single bit. We also used periods of 3, 10 and 30 s of consecutive shots at 10 Hz to write some of the bits. The laser beam was focused onto the sample by using a (0.5, ×20) microscope objective. A (*x-y-z*) translation stage was employed to control the position of the sample. The reference and transmission signals were focused and then collected onto photodiodes D1 and D2, respectively. The writing phase was monitored by measuring the induced decrease in the transmission signal, which indicates non-linear interaction of the laser beam with the medium.

**c) Reading**

Photo-induced variations are read in a linear low flux regime. The main idea is that the localized induced spots backscatter incident light due to the refractive index variation. Reading the spot from a particular layer is achieved with a full-field OCT device[25] as shown in figure 2. This setup is a modified Michelson interferometer where each arm contains the same microscope objective (Linnik Configuration), in order to image the sample and a reference mirror. The optical signal that we measure is the interference pattern between the backscattered layer and the reference beam. For this purpose, we use a DALSA system CCD camera -256×256 pixels 8 bits- working at a maximum rate of $f$=200 frames per second. The coherence length of the light source (a 100 W tungsten-halogen lamp) is short; consequently, we can obtain images with an axial resolution ($z$) close to 1 μm. The lateral resolution ($x,y$) of the setup depends upon the numerical aperture of the objectives. We commonly use (0.25, ×10) and (0.40, ×20) objectives reaching ~1 μm of lateral resolution. The acquisition time is, in general, of 1 s.

The interference pattern is sinusoidally modulated by the vibration of the reference mirror induced with a piezoelectric transducer. This modulation rates at $f/4$=50 Hz, and it is synchronized to the image acquisition of the camera by means of two phase-locked generators. As fully described elsewhere[26], the pixels of the camera are treated in parallel; each image within the sample (corresponding to a path retardation) is the record of a sequence of four acquisitions. The amplitude and phase of the signal are given by linear combinations between these four shifted images. This technique allows a fast parallel-reading scheme, with a very high axial resolution.

## 3. Results and Discussion

The bits (data marks) were recorded in two sol-gel samples. Undoped sol-gel and Perylene orange-doped sol-gel matrices were submitted to single laser pulses between 0.1 and 100 J/cm$^2$. The number of pulses was tuned as indicated in the experimental section, from 1 to 32 single-shots, and from 3 to 30 s of consecutive shots at 10 Hz. This control of the incident energy is shown in figure 3 for the Perylene orange sol-gel matrix: the minimum energy density necessary for writing into the Perylene orange sample was of 10 J/cm$^2$ with 4 single-shots per bit. There is no trace of recording for lower energies. For the undoped sample, this minimum was of 33 J/cm$^2$ with one single-shot per bit. As it is also seen in the figure, single-shot recording at 33 J/cm$^2$ is possible for the Perylene orange sample, indicating the feasibility of systematic single-shot recording with our system. The recording at 100

J/cm$^2$ was detected in a different plane than that shown in the figure. This is due to an effect of self-focusing of the beam, as it will be discussed below.

As it is shown in figures 4(a) and 4(b), by using the energy density of 33 J/cm$^2$, with 2, 4 and 8 laser pulses per bit (left to right), we recorded two different data pages into the Perylene orange bulk sample. A distance of 60 μm between the page planes was measured, with the first of them located 115 μm under the surface. The distance between bits was of 60 μm. Figure 5 shows also the readout signal through a row of 5 bits recorded in the same sample by using the minimum energy density of 10 J/cm$^2$ with 4 laser pulses per bit. They were separated by a distance of 20 μm, each 5-7 μm in diameter, with an axial length of ~10 μm. The power-signal to noise ratio was found to be around 140 for an acquisition time of 1 s. This ratio is rather large and may allow a faster acquisition, for example within 100 ms with a resultant power-signal to noise ratio of 14. These results demonstrate the potential of using dye-doped sol-gels as a medium for 3D optical data storage as well as our OCT-based parallel-reading scheme.

We compared the level of the backscattered signal from the data marks to the one obtained from the surface of the sample. Considering a reflection coefficient of 4% for the surface, we deduced an equivalent reflection coefficient of $R \approx 2.5 \times 10^{-7}$ for the data marks. For the sake of comparison, if we were to assume an uniform effect over the depth of the data marks, this equivalent reflection coefficient would correspond, using a classical Fresnel approximation, to an induced refraction index variation of $\Delta n \approx 8 \times 10^{-4}$ for the recorded bits. While increased $\Delta n$ will also increase the backscattered signal, one must keep in mind that the signal read from "lower" pages must propagate through the "upper" pages of randomly distributed data marks, which will scatter the signal, increasing the cross talk and affecting the device performance. Moreover, an increased refraction index change may arise from a larger induced change into the matrix, which may imply spending more energy in recording, and therefore a more expensive writing system. Additionally, according to this evaluation, the sensibility of our new reading device is high enough for the readout of the recorded information in our system.

Our initial goal was to photo-bleach the molecules doping the material by using TPA, but the fact that bit recording is possible into an undoped sol-gel matrix demonstrates the presence of other processes. Furthermore, as it is shown in figure 6 for the Perylene orange sample, within its optical limiting curve, the material optical damage and TPA energy thresholds are very close. This result was also observed for other differently doped sol-gel materials, notably when doping with Rhodamine and Pyrromethene families. It was even observed for a conventional $SiO_2$ glass sample.

In order to explain these results, multi-photon, thermal effects and dielectric breakdown must be taken into account. Other results in the literature show the possibility of recording in transparent materials by generating ultrafast-laser micro-explosions[16, 27]. In this case, the energy deposition is initiated by multi-photon absorption[28] and may lead to stationary thermal lensing[29], which may also initiate a strong self-focusing of the beam[28, 30], even if it is already externally and tightly focused. As a consequence of this self-focusing, a hot, high-density electron plasma is produced[30]. This plasma has been effectively observed (a white flash inside of the sample occurring at the moment when the pulse is delivered) in both of our samples and as a direct effect, it induces a refractive index change. According to the results found by other groups, the plasma transfers its excess energy to the lattice by electron-phonon coupling[28, 31]. This transfer has a lifetime ranging between 10 and 70 ps that is strongly dependent on the pulse energy[30]. When femtosecond (fs) pulses were used, this self-focusing initiated very small heat affected zones in the materials, and so very small sized recorded bits[28, 31]. However, in picosecond regime, where plasma energy transfer occurs during the pulse application itself, the resulting structures are larger, irregularly shaped, and display radially extending cracks[27]. This explains the irregularity of the backscattered readout signal over the profile of the two lines of bits showed in figure 3. This profile is seen from the lateral dotted line point of view in the *x*-axis, as we can see in figure 7.

Otherwise, it has been suggested that the measured index variation implies that the micro-explosions create voids in the material, and then a compaction of the surrounding region[28]. Additionally, the actual bit spot size is very sensitive to the spatial mode of irradiation pulses, especially for energy into these limits[15]. Finally, the threshold difference for the energy densities observed in the recording of the pure and doped samples is an evidence of a molecular effect. This may imply a pure molecular TPA besides a matrix multi-photon absorption leading to a micro-explosion recording, as mentioned above. Therefore, additional experiments are being performed with our system in order to clarify the recording mechanism into undoped and doped sol-gel materials. These experiments include the use of better chromophores with higher TPA-cross sections, molecular concentration control by binding the molecules to the sol-gel matrix, spatial filtering of the laser beam and fs-recording.

## 4. Conclusions

Data recording and reading have been performed for the first time into undoped and doped sol-gel materials. Recording used 1064 nm picosecond optical pulses. The energy threshold was found to be 33 and 10 $J/cm^2$ for undoped and perylene-doped sol-gel samples, respectively. A new efficient parallel-reading system based on a full-field Optical Coherence Tomography-technique has been proposed. We have shown the possibility of 3D optical storage in sol-gel materials with a reading in parallel. The experimental results show the potentiality of our system for this kind of application, although further research is needed in order to better understand the recording mechanism and to improve it.

## 5. Acknowledgements

This work has been supported by a grant coming for the ACI-Photonics project from the Ministère de la Recherche (France). J. A. Reyes-Esqueda acknowledges Conacyt (Mexico) and the Ministère de la Recherche (France) for their support for a postdoctoral scholarship. L. Vabre acknowledges the Délégation Générale pour l'Armement (France) for its support.

# Figure Captions

**Figure 1**. Experimental system for writing information in 3D. BS: beam splitter. L: lens. ND: neutral optical density. O: microscope objective. D1, D2: photodiodes.

**Figure 2**. Schematic representation of our Linnik interference microscope. A piezoelectric transducer induces a phase modulation (50 Hz), which is synchronized to the image acquisition of a CCD camera (200 Hz) by means of two generators.

**Figure 3**. Control of the incident energy for the Perylene orange sample. For a field of view of 600×600 μm$^2$, two lines of bits written at 33 and 10 J/cm$^2$ are shown (upper and lower, respectively). The distance between bits is of 40 μm. Each bit was marked with a different number of pulses, varying from 1 single-shot to 30 s of consecutive shots (from right to left). The minimum energy density necessary to have a bit-mark was of 10 J/cm$^2$ with 4 single-shots per bit (the encircled bit). See figure 7 for the dotted line.

**Figure 4**. (a) and (b) show two pages of data separated by 60 μm in the axial direction. The field of view is 360×360 μm$^2$ for an exposure time of 1 s and an (0.25, ×10) objective. The distance between bits is of 60 μm.

**Figure 5**. The readout signal is shown through a row of 5 bits recorded on a plane in the Perylene orange sample at 10 J/cm$^2$ with four laser pulses per bit. The field of view is 200×200 μm$^2$ for an (0.40, ×20) objective. The distance between bits is of 20 μm.

**Figure 6**. Limiting curve of the perylene orange sample showing the superposition of its damage material and TPA thresholds. The non-linear coefficient $\beta$ was calculated by fitting the experimental data to the normalized transmission when TPA occurs for a gaussian beam.

**Figure 7**. Transversal view of figure 3 from the lateral dotted line point of view in the *x*-axis and integrated between the two white lines on there. The irregularity of the backscattered readout signal can be seen over the profile of the two lines of bits.

Figure 1.

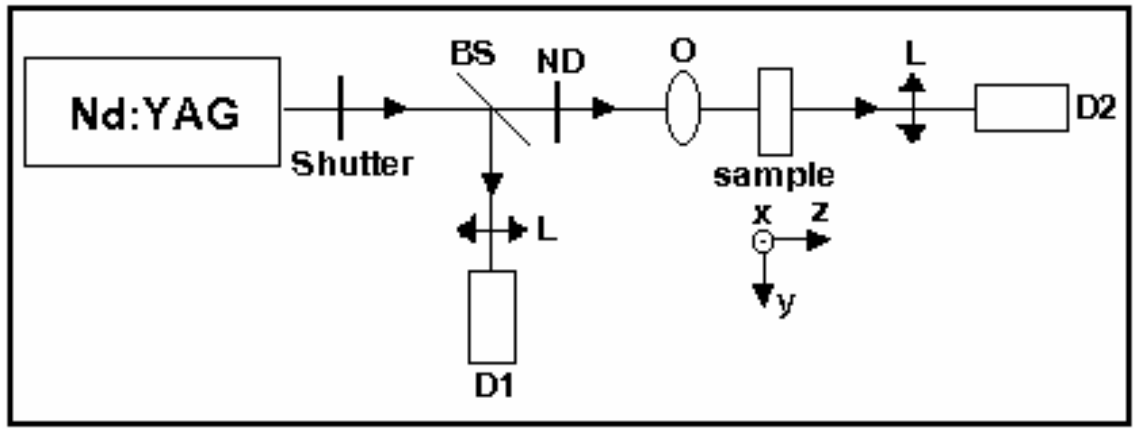

Figure 2.

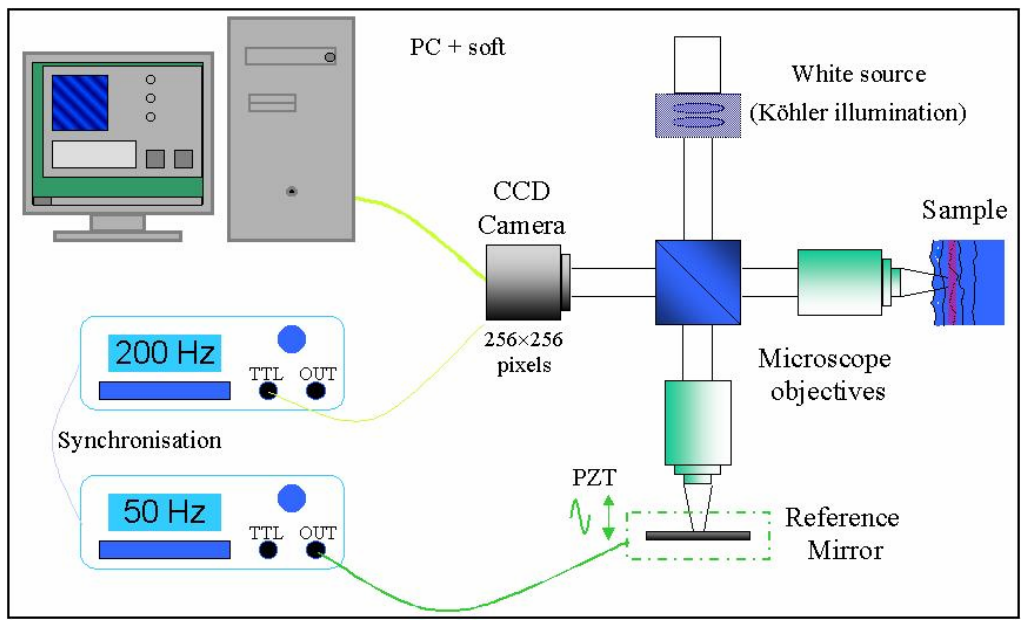

Figure 3.

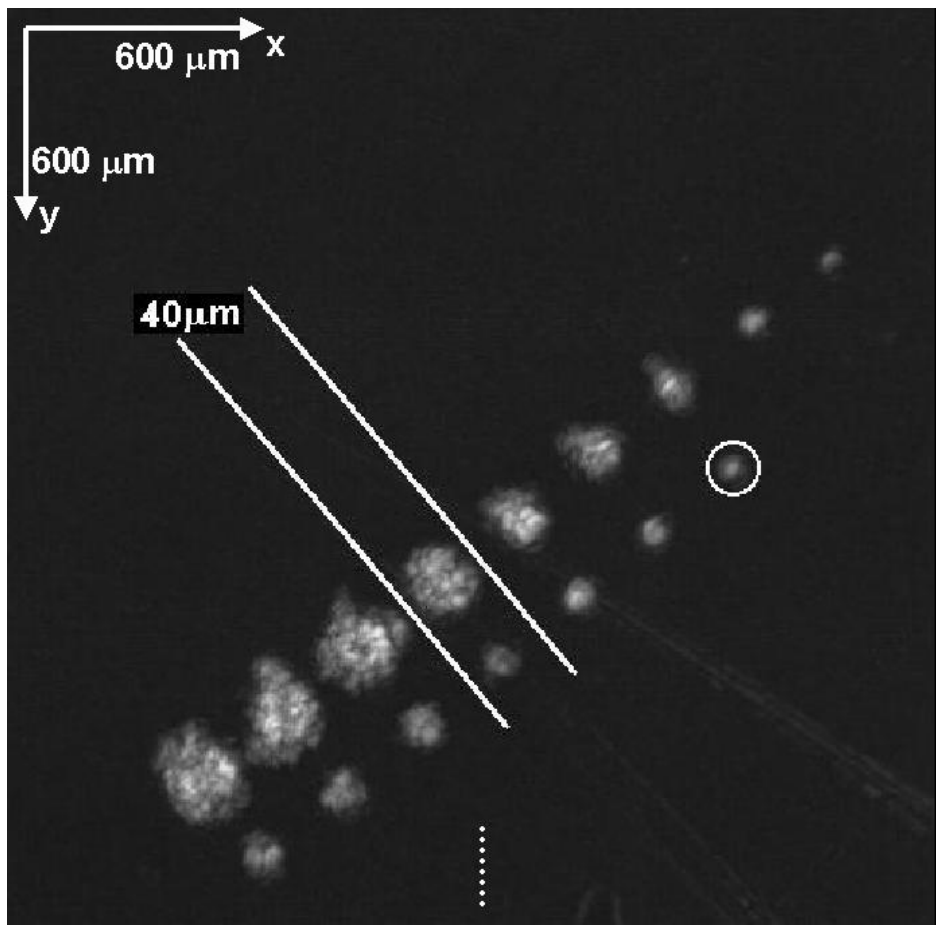

Figure 4.

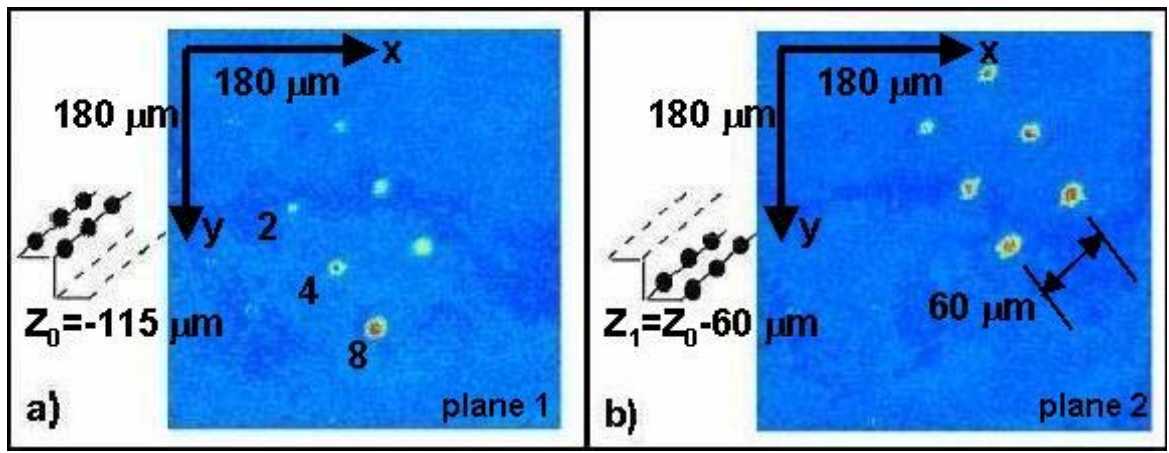

Figure 5.

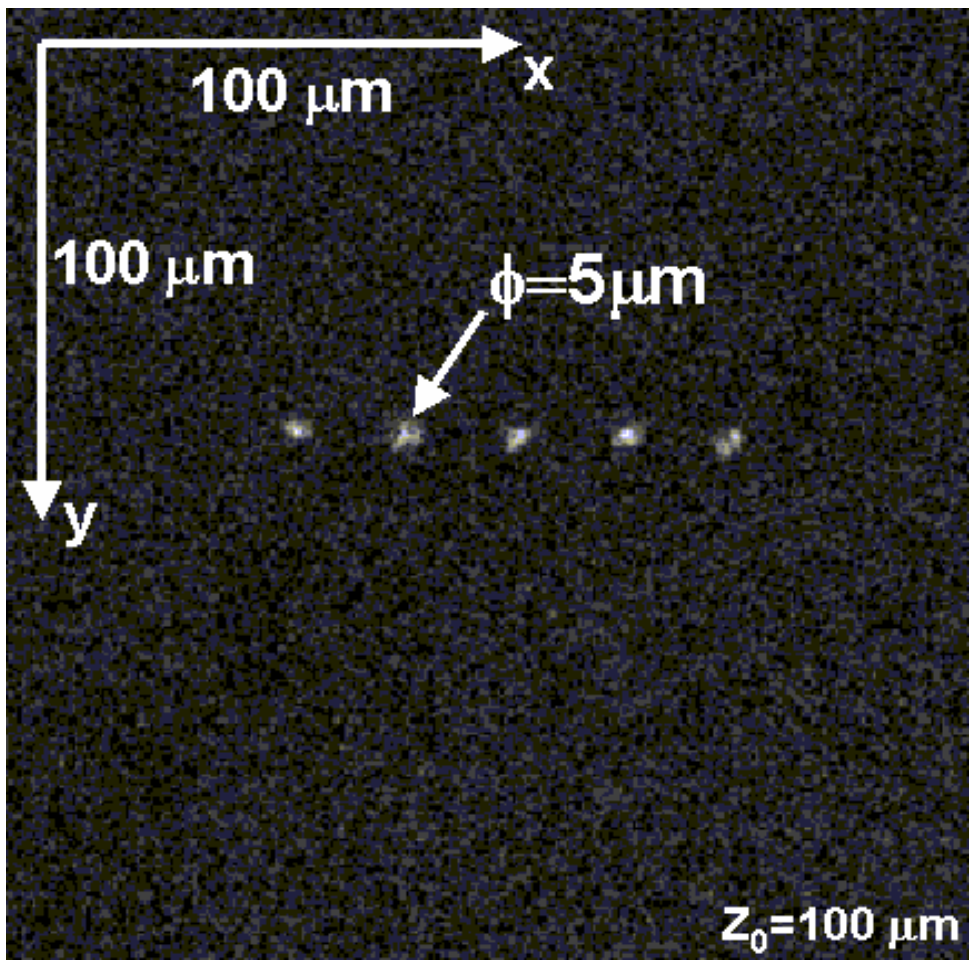

Figure 6.

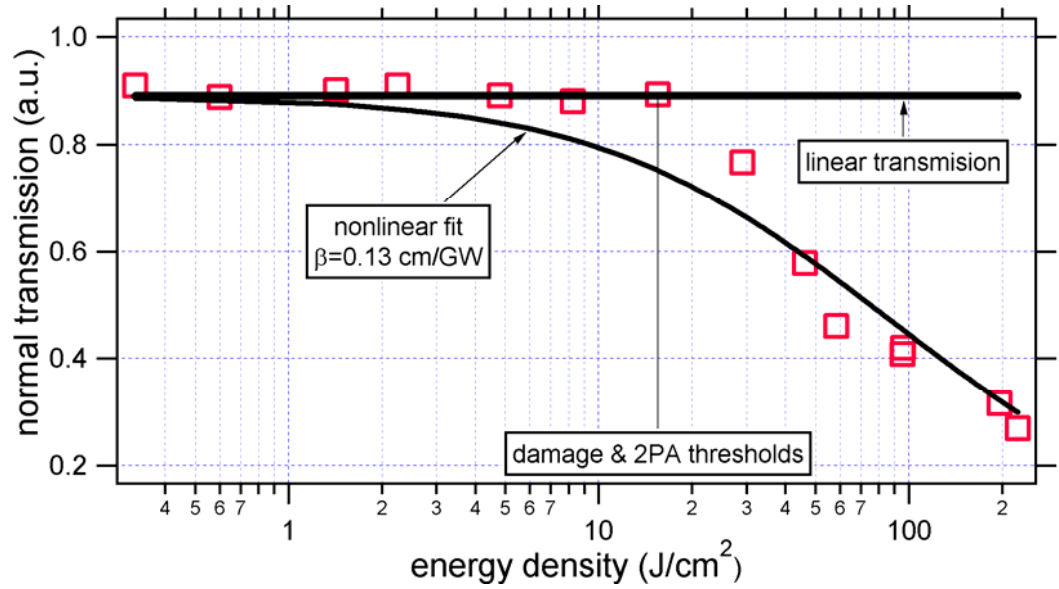

Figure 7.

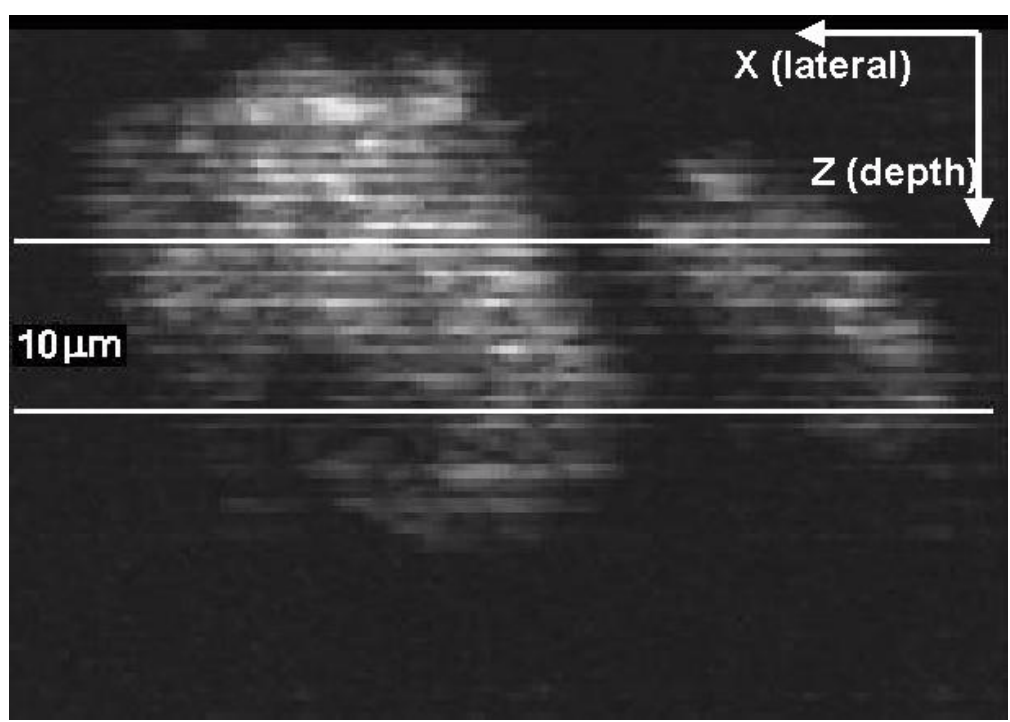